\documentclass[12pt]{article}

\usepackage[letterpaper,margin=3cm]{geometry}

\usepackage{authblk}

\usepackage[parfill]{parskip}
\usepackage{setspace}

\usepackage{sectsty}
\subsectionfont{\fontsize{13}{15}\selectfont}

\usepackage[labelfont=bf,font={small,stretch=0.95}]{caption}

\usepackage[super,sort,compress,numbers]{natbib}

\usepackage{gensymb}
\usepackage{amsmath}
\usepackage{amssymb}
\usepackage{mathtools}
\usepackage{comment}
\usepackage{hyperref}
\usepackage{float}

\usepackage{graphicx}
\usepackage{multirow}

\usepackage[version=4]{mhchem}

\usepackage{mathptmx}

\usepackage{xcolor}
\usepackage{soul}
\usepackage{booktabs}

\newcommand{\etal}{{\em et al.}}

\makeatletter
\renewcommand*{\@fnsymbol}[1]{\ensuremath{\ifcase#1\or\dagger\or*\fi}}
\makeatother

\begin{document}

\title{\Large\bf Distillation of atomistic foundation models across architectures and chemical domains}

\author[1]{John L. A. Gardner\thanks{These authors contributed equally.}}
\author[1]{Daniel F. Thomas du Toit$^{\dagger}$}

\author[1]{Chiheb Ben Mahmoud}
\author[1]{Zo\'e{} Faure Beaulieu}
\author[2]{Veronika Juraskova}
\author[1]{Laura-Bianca Pa\c{s}ca}
\author[1]{Louise~A.~M.~Rosset}

\author[2]{Fernanda Duarte}
\author[3,4]{Fausto Martelli}
\author[5,6]{Chris~J.~Pickard}

\author[1]{Volker L. Deringer\thanks{Corresponding author. E-mail: volker.deringer@chem.ox.ac.uk}}

\affil[1]{Inorganic Chemistry Laboratory, Department of Chemistry, University of Oxford, Oxford, UK}
\affil[2]{Chemistry Research Laboratory, Department of Chemistry, University of Oxford, Oxford, UK}
\affil[3]{IBM Research Europe, Hartree Centre, Daresbury, UK}
\affil[4]{CNR-Institute of Complex Systems, Department of Physics, Sapienza University of Rome, P.le Aldo Moro, I-00185, Roma, Italy}
\affil[5]{Department of Materials Science and Metallurgy, University of Cambridge, Cambridge, UK}
\affil[6]{Advanced Institute for Materials Research, Tohoku University, Sendai, Japan}

\date{}

\maketitle

\setstretch{1.1}

{\bf
Machine-learned interatomic potentials have transformed computational research in the physical sciences.
Recent atomistic `foundation' models have changed the field yet again: trained on many different chemical elements and domains, these potentials are widely applicable, but comparably slow and resource-intensive to run.
Here we show how distillation via synthetic data can be used to cheaply transfer knowledge from atomistic foundation models to a range of different architectures, unlocking much smaller, more efficient potentials.
We demonstrate speed-ups of $> 10\times$ by distilling from one graph-network architecture into another, and $> 100\times$ by leveraging the atomic cluster expansion framework. 
We showcase applicability across chemical and materials domains: from liquid water to hydrogen under extreme conditions; from porous silica and a hybrid halide perovskite solar-cell material to modelling organic reactions.
Our work shows how distillation can support the routine and computationally efficient use of current and future atomistic foundation models in real-world scientific research.
}

\clearpage
\setstretch{1.5}

The term `foundation model' (FM) is used to describe a machine-learning model trained on a large, diverse, and therefore general dataset, which demonstrates high accuracy and generalisation behaviour out-of-the-box.\cite{Foundation-Models-21} 
Such models are currently transforming research and technologies in many fields, including natural language processing, computer vision, medicine, and the physical sciences.\cite{Brown-20-07,Kirillov-23-04,zhou_foundation_2023,Parker-24-07,Hollmann-25-01, fu_foundation_2025, pyzer-knapp_foundation_2025, bodnar_foundation_2025}
For example, an image-recognition model was trained on large amounts of unlabelled medical images and then fine-tuned (adapted in a supervised setting) to predict the occurrence of diseases; \cite{zhou_foundation_2023} a transformer model was trained on heterogeneous data comprising weather forecasts, ocean-wave and hurricane dynamics to yield a foundation model for the Earth system as a whole. \cite{bodnar_foundation_2025}

In the domain of atomistic simulations of materials and molecular systems, the term `foundation model' is typically used more specifically: for machine-learned interatomic potential (MLIP) models that have been trained on very large datasets of diverse chemical systems.
For solid-state materials, notable examples include MACE-MP-0,\cite{Batatia-24-03} MatterSim,\cite{Yang-24-05} Orb,\cite{Neumann-24-10, Rhodes-25-04} and OMat24;\cite{Barroso-Luque-24-10}
for molecular systems, the ANI and AIMNet model series has long been established,\cite{Smith-17-03,Zubatyuk-19-08} and the MACE-OFF models have been proposed more recently.\cite{Kovacs_JACS_MACEOFF_2025}
There are also emerging models that target both the materials and molecular domains. \cite{Shoghi-24-05, Zhang-24-05, Zhang-24-12}
The accuracy of MLIPs in general, and of atomistic FMs in particular, continues to improve at a rapid pace, including via architectural innovations,\cite{Liao-24-03, Fu-25-02, Batatia-25-01} enhanced benchmarking  \cite{Pota-24-09, Riebesell-24-12} and training protocols,\cite{Shiota-24-12} and improved datasets.\cite{Schmidt-22-03, Barroso-Luque-24-10, Kaplan-25-03,Levine-25-05}

An important next step for MLIPs is now to make them easily accessible and computationally cheap to use --- to the extent that they can be considered genuine mainstream methods, usable by anyone with only modest hardware resources rather than requiring access to top-tier computing centres. 
This challenge is particularly pronounced for atomistic FMs, which often rely on (and derive much of their success from!)\ sophisticated graph-based architectures and large model sizes.
Transferring information from an expensive model to a cheaper one is a  well-known concept in machine-learning research:
initially proposed by Hinton \etal, \cite{Hinton-15-03} \textit{knowledge distillation} refers to any process by which information learned by one model (the teacher) is transferred to another model (the student:\ typically smaller, simpler, and faster), \cite{Hinton-15-03, Allen-Zhu-23-02} and this idea is now widely applied in different research fields.\cite{Cheng-20-03, Touvron-23-02, Peng-23-04, Liu-24-02, Vemulapalli-24-07} 
A popular approach to perform knowledge distillation is via the use of synthetic data: for example, improvements on the foundation model of Ref.~\citenum{zhou_foundation_2023} were possible by pre-training a model on a large amount of synthetic medical images generated with stable diffusion. \cite{Sun-25-03} 
For atomistic ML, the term `synthetic data' refers to structures and labels generated by a surrogate model, rather than the ground-truth quantum-mechanical method. \cite{Gardner-23-03}

Recent work has introduced several initial techniques for distilling MLIPs.
Beyond aiming to reproduce a teacher model's total-energy and force predictions,\cite{Morrow-22, Zhang-24-12,Rhodes-25-04}
other approaches focus on the teacher's {\em local}-energy predictions, \cite{Gardner-23-03, Matin-25-02}
learn the teacher model's Hessians, \cite{Amin-25-01}
align the internal representations of the teacher and student models, \cite{EkstromKelvinius_ANIPS_Accelerating_2023}
or learn from an ensemble of teachers. \cite{Matin-25-03}
The use of synthetic data for MLIP training was initially explored in the context of fast linear models,\cite{Morrow-22} then of graph-neural-network models,\cite{Gardner-24-01} and more recently for one of the currently proposed atomistic FMs.\cite{Wang-25-02}
However, most work in this space reported to date has focused on specific MLIP architectures, and often used domain-specific data-generation protocols (such as customised MD simulations in Refs.~\citenum{Gardner-23-03} and \citenum{Morrow-22}).

Herein, we propose a general, architecture-agnostic protocol for distilling any atomistic FM to target any user-specified chemical domain of interest (Fig.~\ref{fig:overview}a).
Our approach quickly produces fast, accurate, and stable student MLIPs that can be used to drive extensive simulations with modest computational resources.
Being based on generalised data-driven knowledge transfer, our approach transcends specific MLIP fitting frameworks, and indeed one of its most promising applications is one that does {\em not} require a computationally expensive graph-network architecture for downstream applications. 
Building on previous milestone work that has shown that (i) MLIPs can be trained based on a local, atom-centred decomposition of the total energy; \cite{Behler-07-04, Bartok-10-04} (ii) graph-based MLIPs provide unprecedented accuracy and flexibility for this task; \cite{Batzner_NC_E3equivariant_2022, Chen-22-11, Batatia-23-01} and that (iii) the latter can be used to fit chemically comprehensive models that scale well with data, \cite{GNoME, Deng-23-09, Batatia-24-03, Yang-24-05} we believe that distillation is now the fourth and final major methodological step in MLIP development, unlocking routine and efficient applications for mainstream materials and molecular modelling.

\section*{Proof-of-concept}

\begin{figure}[ht]
    \centering
    \includegraphics[width=\linewidth]{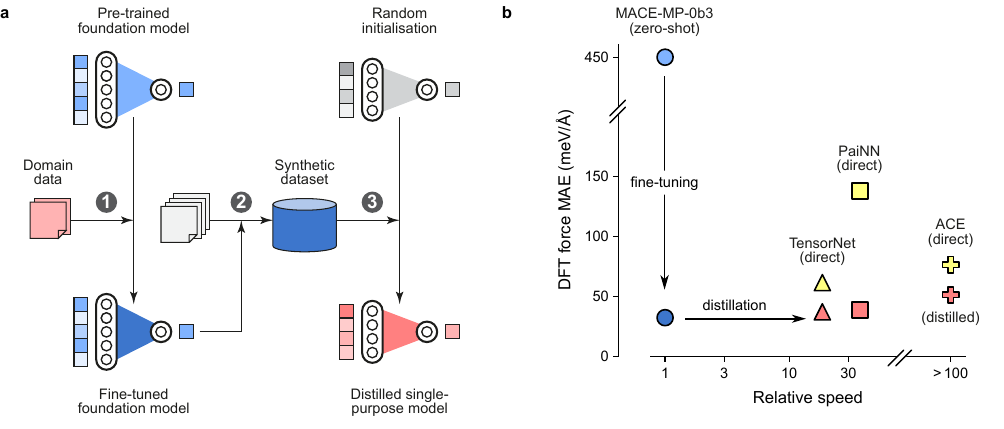}
    \caption{
    \textbf{Distilling atomistic foundation models via synthetic data.} 
    (\textbf{a}) Illustration of the three-step-procedure used throughout the present work. 
    \textbf{1}: An existing foundational MLIP model (\textit{light blue}) is fine-tuned on a small amount of domain-specific structures with quantum-mechanical energy and force labels (\textit{red}).
    \textbf{2}: This fine-tuned `teacher' model (\textit{dark blue}) is then used to cheaply generate and label a much larger dataset of synthetic structures, similar to Ref.~\citenum{Morrow-22} but now without the need for MD simulations (Methods). 
    \textbf{3}: Finally, a small and fast `student' model is trained on this synthetic dataset in a near-automated fashion, yielding the distilled single-purpose MLIP model highlighted in red.
    (\textbf{b}) Proof-of-concept for this protocol for liquid water. 
    Relative to the \texttt{MACE-MP-0b3} foundation model, \cite{Batatia-24-03} distilled models using the \texttt{TensorNet}, \texttt{PaiNN}, and \texttt{ACE} architectures are substantially faster at nearly the same accuracy on the DFT-labelled test set. 
    We include models directly trained on the domain-specific DFT dataset for comparison (\textit{yellow markers}): these have seen only small amounts of data and are uniformly worse than their respective distilled counterparts (\textit{red}).}
    \label{fig:overview}
\end{figure}

As a proof of concept, we distil the \texttt{MACE-MP-0b3} foundation model \cite{Batatia-24-03} to target liquid water at room temperature and pressure, aiming to create much faster student potentials at the hybrid-DFT (revPBE0-D3)\cite{Ernzerhof_JCP_Assessment_1999, Adamo_JCP_Reliable_1999, Grimme_JCP_Consistent_2010, Goerigk_PCCP_Thorough_2011} level of theory.
Figure \ref{fig:overview}a illustrates our protocol.
We begin by splitting an existing, high-quality dataset for bulk liquid water, taken from Ref.~\citenum{Cheng-19-01}, into a training set of 25 structures, a validation set of 5 structures, and a test set of 1,563 structures. We emphasise that the training set is extremely small, consistent with previous work that showed successful fine-tuning of MACE models with only very few examples. \cite{Kaur-25-01}

The first step of the approach is the fine-tuning of the FM. 
\texttt{MACE-MP-0b3} was trained using a mid-range (PBE+U) level of DFT: fine-tuning therefore involves raising the level of the model and drastically improves accuracy on the test set (component-wise force \textit{R}$^2$: 0.944 $\rightarrow$ 0.9991, component-wise force mean absolute error, MAE: 450 meV/\AA\ $\rightarrow$ 32 meV/\AA). We focus on \texttt{MACE-MP-0b3} as one of the currently popular FMs for the present proof-of-concept, but we note that our protocol illustrated in Fig.~\ref{fig:overview}a is \textit{architecture agnostic}, and in the Applications section we include representative examples of distilling other FMs as well.

The second step is synthetic data generation.
We use the fine-tuned FM to augment the initial 25+5 training+validation structures via the iterative sequence of rattling and relaxation steps that is detailed in the Methods section. 
This protocol is extremely sample-efficient, requiring an average of five model calls to produce each new un-labelled, synthetic structure --- at least an order of magnitude more efficient than MD.
We then use the fine-tuned FM (not DFT!) to label the resulting $\approx$ 10k structures, as well as 50 structures for each H--H, O--O and O--H dimer.

\begin{figure}[t]
    \centering
    \includegraphics[width=0.6\linewidth]{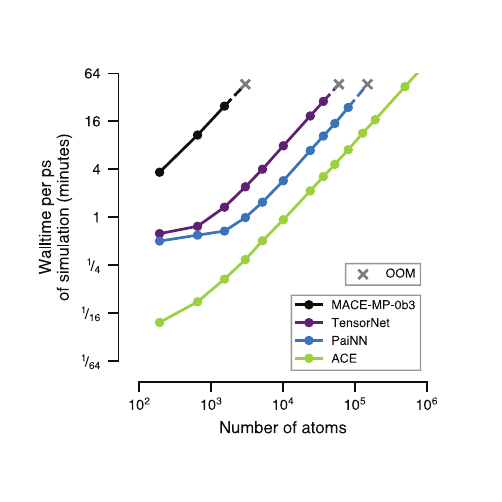} 
    \caption{\textbf{Computational efficiency.}
    We show the inference costs for different water models. Speeds are measured during 300~K NVT simulations run on a single Nvidia RTX A6000 GPU. The (fine-tuned) \texttt{MACE-MP-0b3} model is memory-intensive, and quickly runs out of memory (OOM) as the system size increases beyond 1,000 atoms. In contrast, the smaller, distilled GNN models allow us to scale up system size by up to another 2 orders of magnitude. 
    Finally, ACE provides another order of magnitude in speed-up, and allows for scaling to very large systems (OOM at $\sim6.7\times10^{7}$ atoms).}
    \label{fig:timings}
\end{figure}

The third and final step is to train a variety of small and fast student MLIPs on these synthetic structures and labels, making use of the increased computational speed compared to the FM teacher model.
We arrive at \texttt{TensorNet}, \texttt{PaiNN}, and \texttt{ACE} models with force errors relative to DFT that are very close to that of the fine-tuned FM (component-wise force MAEs of 37, 39, and 51~meV/\AA\ respectively); see Fig.~\ref{fig:overview}b.
These distilled models are more than $10 \times$ faster than the FM and improve in accuracy over direct training of the respective MLIP architectures (on the starting dataset) by up to 70\%.
Crucially, the distilled models are capable of driving MD simulations on a single GPU that are stable, unlike the directly-trained models, and can be scaled to large system sizes, unlike the expensive FM (Fig.~\ref{fig:timings}).

Compared to \texttt{PFD} --- a recently published package for fine-tuning of DPA FMs and distillation into DeepMD models \cite{Wang-25-02} --- our method differs by offering architectural flexibility for users: compatible with any ASE calculator object, \texttt{augment-atoms} can be used with a variety of FMs, and we distil into multiple model architectures, as they can be fit with any fitting software once the synthetic dataset is generated.
The generation of the synthetic dataset is reduced in cost by sampling without MD, allowing the fitting of accurate distilled models without the need for server-grade hardware.

In total, our approach took 7.5 hours to go from starting FM to distilled \texttt{PaiNN} model on a single, mid-level GPU (Nvidia RTX A6000): approximately 30 minutes to fine-tune the foundation model, 180 minutes to generate and label 10k synthetic structures, and 230 minutes to train the distilled model. Hence, with a comparably small investment of resource and energy, and using only a single GPU, the user can speed up the simulation as compared to the expensive foundation model.

\begin{figure}[H]
    \centering
    \includegraphics[width=\linewidth]{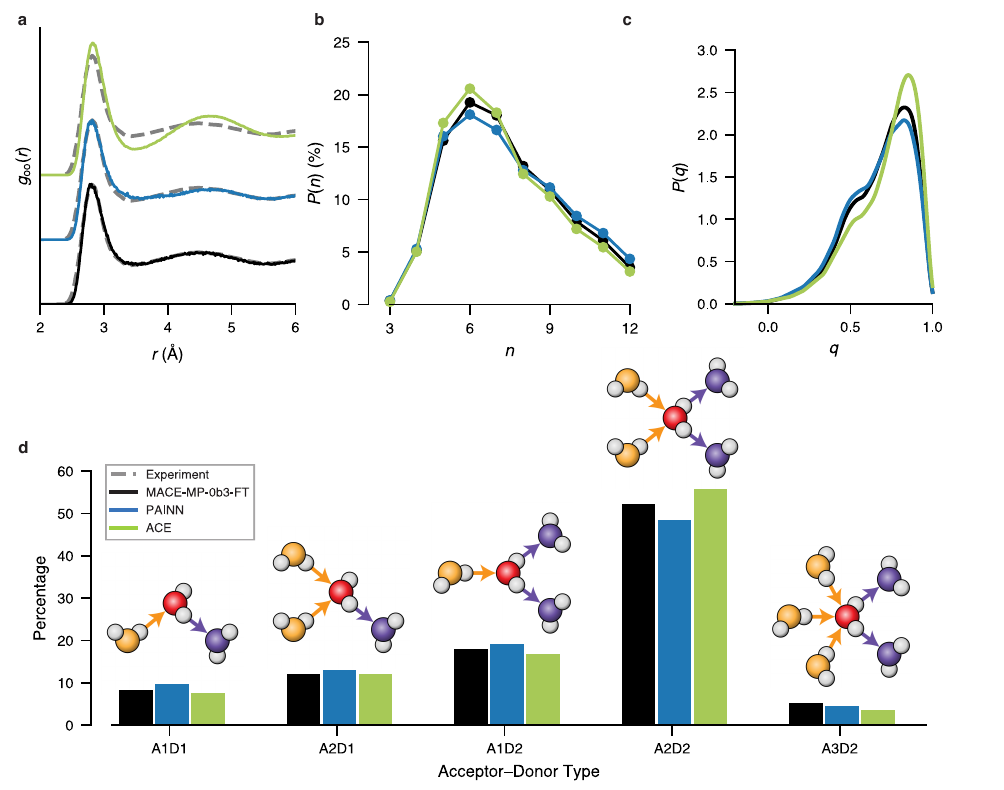}
    \caption{\textbf{Physics-guided validation of distilled MLIPs.}
    All analyses are derived from MD simulations of liquid water at 300 K.
    (\textbf{a}) Radial distribution function for O$\cdots$O contacts. The distilled PaiNN model almost perfectly reproduces the behaviour of the FM, which in turn agrees well with experimental data (taken from Ref.~\citenum{skinner_structure_2014}). 
    (\textbf{b}) Distribution of $n$-membered rings in the simulated liquid. 
    (\textbf{c}) Distribution of the tetrahedral order parameter, $q$, \cite{Errington-01-1} in each structural model.
    (\textbf{d}) Hydrogen-bonding network connectivity as in Ref.~\citenum{DiStasio-14-8}, evaluated by counting different categories of topological connectivity, based on the number of H-bond acceptor (A) and donor (D) interactions.
    } 
    \label{fig:water-valid}
\end{figure}

Beyond numerical validation, further tests are required to ascertain the usefulness of an MLIP model. \cite{Morrow-23-03, Fu-25-02} 
In the present case, our quality criterion is how the distilled models describe liquid water in MD simulations.
With all three models, these simulations lead to sensible local structure in terms of the nearest-neighbour environment, indicated by the radial distribution function (Fig.~\ref{fig:water-valid}a).
Consistent with the respective speed-ups, the PaiNN model ($\sim 10\times$ vs the teacher) yields a very similar local structure to experiment and the teacher; the ACE model ($\sim 100\times$) behaves slightly differently and leads to a water structure that appears somewhat over-structured: the first peak in the RDF, indicating O$\cdots$O contacts, is more pronounced compared to experiment.

More insights can be gained from the ring-size distribution, obtained for all rings up to a length of $n = 12$, indicating the medium-range ordering of the liquid (Fig.~\ref{fig:water-valid}b), and from the tetrahedral order parameter of Ref.~\citenum{Errington-01-1} for local environments (Fig.~\ref{fig:water-valid}c). Consistently between all three panels, the atomic environments are less (more) ordered with the PaiNN (ACE) student models, respectively, as seen from the atoms' predicted tetrahedral-like-ness, and from the higher percentage of 6-membered rings (in ideal crystalline ice I, we would have $q=1$ and exclusively six-membered rings).

We finally inspect the hydrogen-bonding (HB) connectivity, which allows for yet more nuanced insight into the simulated structures and therefore for more careful validation. In this analysis, we count the number of hydrogen-bond acceptor (A) and donor (D) interactions for each water molecule, and classify the molecules accordingly, following Ref.~\citenum{DiStasio-14-8}. In crystalline ice I, each water molecule has the same `A2D2' HB topology, whereas liquid water contains many molecules with fewer HBs, and some with more. The descriptions of \texttt{MACE-MP-0b3} and its student models agree well across the range of topologies investigated (Fig.~\ref{fig:water-valid}d); the PaiNN model shows slightly fewer crystal-like environments, in line with the less pronounced structuring evident from the plots in Fig.~\ref{fig:water-valid}a--c; conversely, the ACE model leads to a slightly more ordered (A2D2-rich) water structure.

\section*{Ablation studies}

We now perform ablation studies for several key components and hyperparameters of our protocol: making the training protocol or the student models simpler and testing how doing so affects performance. We use those numerical experiments to study scaling laws for prediction errors, aiming to gain further insight into the protocol in general.

\begin{figure}[H]
    \centering
    \includegraphics[width=\linewidth]{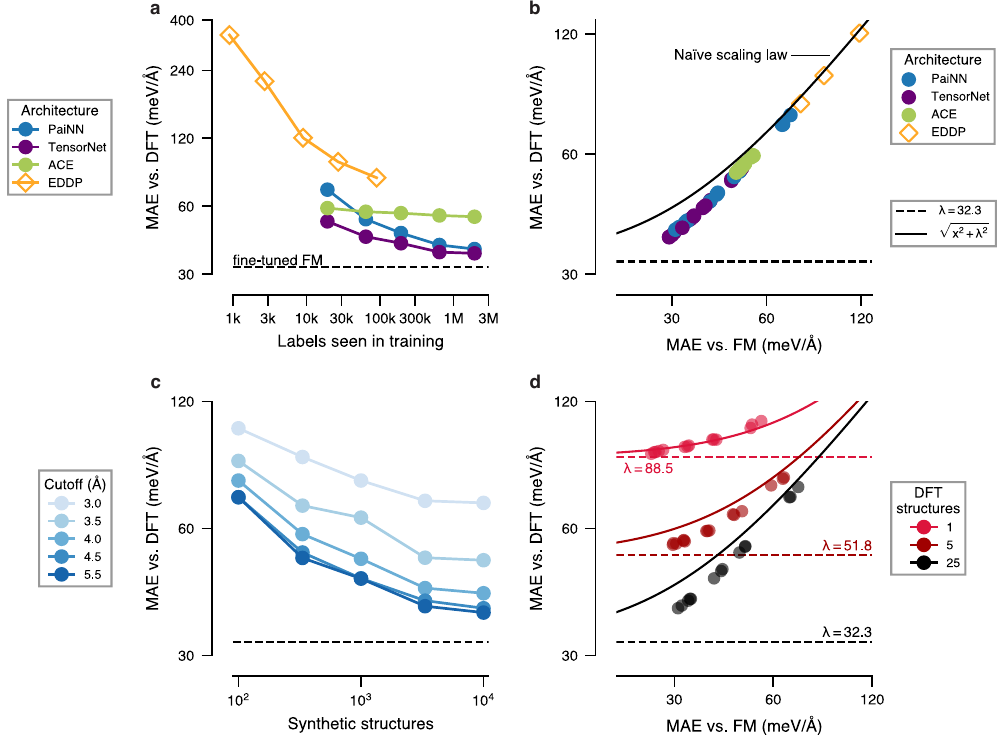}
    \caption{\textbf{Ablation studies.}
    All panels relate to the \texttt{MACE-MP-0b3} foundation model (FM) fine-tuned on water and distilled into various MLIP architectures (see Proof-of-concept section). 
    (\textbf{a}) Learning curves. We test the number of synthetic data labels required for distillation: energy data for small cells in the case of EDDP training ({\em open diamonds}); energy and force data for larger cells in the case of all other models ({\em filled circles}). We plot the distilled models' component-wise force MAE on the DFT-labelled test set while varying the amount of synthetic data used for distillation. All points show the best-of-three. PaiNN and TensorNet models use a radial cut-off of 4.5 \AA{}, while ACE and EDDP use 5.5\AA.
    (\textbf{b}) Scaling laws. We plot the accuracy of each model relative to the fine-tuned FM ($x$-axis) and DFT ($y$-axis) on the hold-out test set. A solid line indicates the distribution we would expect if these two accuracies were uncorrelated; a dashed line shows the error of the FM against DFT (denoted $\lambda$). While the plot is truncated at 130 meV/\AA, the remaining EDDP models behave similarly up to 400 meV/\AA.
    (\textbf{c}) Cut-off radius. We take the same fine-tuned FM as in panel (a), and distil into the \texttt{PaiNN} architecture using various radial cut-offs. Again, all points show the best-of-three.
    (\textbf{d}) Amount of ground-truth domain data (`$n$-shot' performance). We vary the amount of data used to fine-tune \texttt{MACE-MP-0b3} (and hence vary its error against DFT), and plot the MAE of \texttt{PaiNN} models distilled from these fine-tuned FMs relative to the FM itself ($x$-axis) and to DFT ($y$-axis).
    }
    \label{fig:ablations}
\end{figure}

We start by ablating (reducing) the number of synthetic structures used to distil the fine-tuned \texttt{MACE-MP-0b3} model from the previous section (Fig.~\ref{fig:ablations}a).
In addition to the architectures mentioned above, we include ephemeral data-derived potential \cite{Pickard-22-07} (EDDP) models in this series of experiments,  noting that these are trained in a different way (Methods) but are included for a more rounded picture of different architectures.
For all architectures, increasing the amount of synthetic data improves the student models' performance on the DFT-labelled test set, which appears to be approaching the accuracy of the FM in the large-data limit (dashed line). 
The increase in amount of synthetic data comes with a time cost: for this specific combination of system and foundation model, we are able to generate a new, uncorrelated structure once every second, and so data generation for models for 100, 1k, and 10k structures takes 2 minutes, 20 minutes, and 3 hours, respectively.

Our distilled models are trained to mimic the foundation model's outputs on the synthetic dataset: they have never been directly trained on any DFT labels.
In Fig.~\ref{fig:ablations}b, we investigate the relationship between the distilled models' accuracy relative to the FM predictions (x-axis) and the DFT ground truth (y-axis).
For reference, we also plot the `scaling law' we would obtain if these two errors were uncorrelated, viz.\ $y = \sqrt{x^2 + \lambda^2}$, where $\lambda$ is the MAE of the FM on the DFT labels.
For all architectures and amounts of synthetic data, we find that the distilled models are {\em more accurate} on the DFT data than we would expect.
Indeed, there is a negative correlation between the quantities $F_{\text{FM}} - F_{\text{DFT}}$ and $F_{\text{distilled}} - F_{\text{FM}}$: that is, when the FM over-predicts the DFT force, the distilled model tends to under-predict the FM, and hence makes predictions that are closer to the DFT data.
This inherent bias towards the true potential-energy surface (PES) improves the data efficiency of synthetic distillation. A more systematic study is required to understand the origins of this phenomenon.

We carry out two further ablation studies. To be generally applicable across different chemical systems, foundation models need to use large message-passing cut-off radii (with a typical value of 6\,\AA \cite{Rhodes-25-04, Batatia-24-03}).
The number of directly connected atomic neighbours within the cut-off radius $r$ scales as $O(r^3)$: large cut-offs therefore dramatically increase the computational cost of model forward passes.
We show in Fig.~\ref{fig:ablations}c that student models for water can use smaller cutoffs than the FM from which they are distilled, while maintaining good accuracy: in this case, PaiNN student-model predictions using 4.5 and 5.5 \AA{} cut-off radii yield very similar results.

Finally, we ablate the amount of DFT-labelled structures used to fine-tune \texttt{MACE-MP-0b3} (Fig.~\ref{fig:ablations}d).
More fine-tuning data lead to more accurate foundation models, which in turn leads to more accurate distilled models with respect to the DFT ground truth. Such fine-tuning in the very-low-data regime could be particularly interesting for cases where very-high-level (beyond-DFT) ground-truth data are to be used, as pointed out previously. \cite{Gardner-23-03}

\section*{Applications}

In this section, we describe example applications of our distillation approach to representative systems in materials science and chemistry. We aim to demonstrate the general validity of the approach using relatively simple tests; we expect that future studies will apply the distilled potentials (or similar ones) to downstream research questions.

\subsection*{Metallic hydrogen}

Our first test case is hydrogen, the lightest and most abundant chemical element in the universe, at extreme temperature ($> 10,000$~K) and pressure ($> 400$~GPa), following Ref.~\citenum{BenMahmoud-22-09}. In this regime, hydrogen exists as a dense metallic liquid, resulting from the dissociation of \ce{H2} molecules. Such conditions are thought to exist in the core of giant planets or brown dwarf stars.\cite{Guillot_AREPS_Interiors_2005, Helled_NRP_Understanding_2020} Experiments at these extreme conditions remain challenging, \cite{Celliers_S_Insulatormetal_2018} and so quantum-mechanically based simulations are important for exploring the element's equation of state (EOS). Establishing the EOS requires running MD simulations with many individual (very) short timesteps to obtain accurate estimates of densities and phase stabilities. 

We fine-tuned \texttt{MatterSim-v1.0.0-1M} on 100 liquid hydrogen configurations at densities up to 1.6 g cm$^{-3}$, taken from Ref.~\citenum{BenMahmoud-22-09}, before distilling into a 40k parameter \texttt{PaiNN} model with a 2 \AA{} radial cutoff using 5k synthetic structures. The reduction in model size (96\% reduced parameter count, 15$\times$ reduction in the volume of the receptive field) compared to the FM makes the distilled model more than 2 orders of magnitude faster at test time. We validate our distilled model by generating pressure--density equations of state (`isotherms') at temperatures between 10,000 and 50,000 K in Fig.~\ref{fig:applications}a, and compare to first-principles MD isotherms as reported in Ref.~\citenum{BenMahmoud-22-09}. We note that in this temperature / pressure regime, it is important to incorporate finite electronic temperature effects:\cite{Karasiev_PRL_Nonempirical_2018,Bonitz_PoP_Initio_2020} for example, discrepancies in predicted volumes between ground-state and finite-temperature computations can be as high as 10\% at 50,000~K.\cite{BenMahmoud-22-09} Nevertheless, even without considering finite temperature effects, as we do in this work, metallic liquid hydrogen is still a challenging benchmark. 

Figure \ref{fig:applications}a shows that the distilled model's EOS agrees near perfectly with DFT within the domain on which it was trained (densities below 1.6 g cm$^{-3}$).
However, when the distilled model is evaluated in an out-of-distribution setting (densities above 1.6 g cm$^{-3}$), its behaviour begins to diverge from the ground-truth. Care should therefore be taken by practitioners to ensure a diverse range of structures are used for FM fine-tuning and initialising the synthetic data generation, such that they fully cover the desired application area.

\subsection*{Porous amorphous silica}

Porous amorphous materials pose a challenge for atomistic modelling, with aspects of both bulk and surface physics at play. Porous amorphous silica (a-SiO$_2$) aerogels are of interest as insulators for space due to their light weight and low thermal conductivity. \cite{gross_mechanical_1988, abdusalamov_modeling_2021,liu_microscopic_2022, iswar_dense_2021} Here we select mesoporous a-SiO$_2$ as another benchmark for our distillation protocol.

We distil a \texttt{PaiNN} model from \texttt{orb-v3} fine-tuned on 25 underdense a-SiO$_2$ structures from Ref.~\citenum{Mahmoud-24-12}. The distilled \texttt{PaiNN} model has `seen' some small pores of up to 1 nm across, but has not been trained explicitly on surfaces.
We test our potential by running constant-pressure MD simulations (10 ps, 500 K, 1 bar) on a mesoporous ($\rho = 1.2$ g/cm$^3$) a-\ce{SiO2} structure of 1,200 atoms, generated using a general-purpose \texttt{ACE} potential for the Si--O system \cite{Erhard-24-03}.

\begin{figure}[H]
    \centering
    \includegraphics[width=\linewidth]{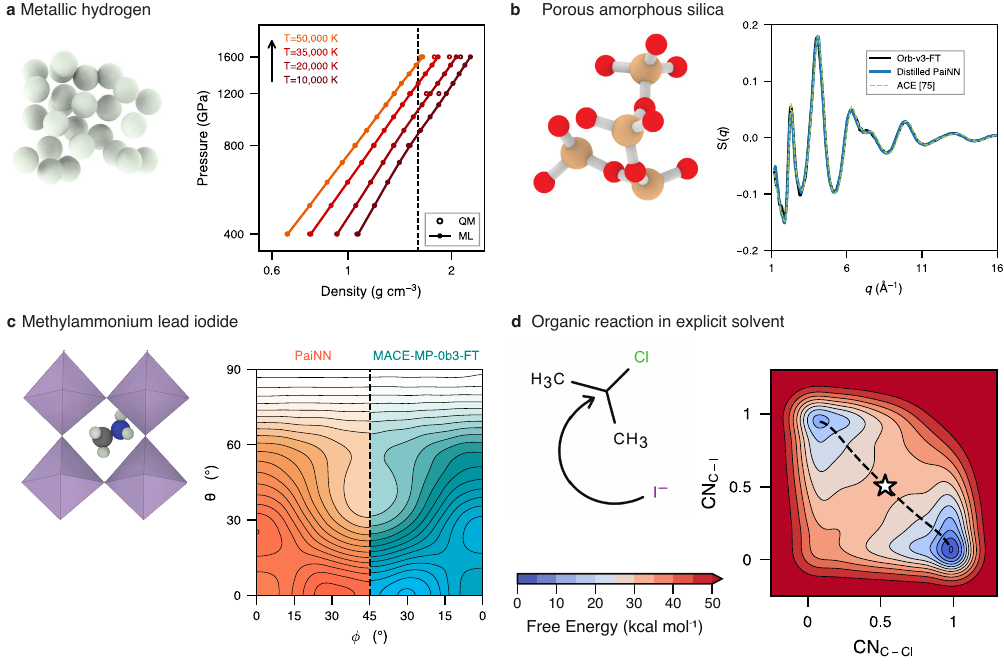}
    \caption{\textbf{Applications of distilled MLIPs across different domains.}
    (\textbf{a}) Metallic hydrogen. We compute pressure--density plots at constant temperatures (`isotherms') and compare to predictions from DFT,\cite{BenMahmoud-22-09} testing applicability under extreme conditions. The dashed line shows the maximum density for which configurations are included in the fine-tuning set.
    (\textbf{b}) Porous amorphous silica. We compute the X-ray structure factor, $S(Q)$, for structural models from MD simulations, and compare against the state-of-the-art MLIP model of Ref.~\citenum{Erhard-24-03}.    (\textbf{c}) Methylammonium lead iodide. We measure the characteristic angles, $\theta$ and $\phi$, defining the orientation of the MA$^{+}$ cation, during MD simulations. The heatmaps show the distribution of angles, comparing the distilled PaiNN model ({\em left}) with its fine-tuned MACE `teacher' ({\em right}).  
    (\textbf{d}) A `textbook' organic reaction in explicit solvent. We use well-tempered metadynamics to predict the free-energy surface for the S$_{\text{N}}$2 substitution reaction between 2-chloropropane and an iodide ion in explicit acetone solvent.
    The structures in panels (a--c) were visualised using OVITO. \cite{ovito}
    }
    \label{fig:applications}
\end{figure}

We compare our \texttt{PaiNN} student model to the \texttt{orb-v3-FT} teacher, and to the \texttt{SiO-ACE-24} reference model \cite{Erhard-24-03} in Fig.~\ref{fig:applications}b, showing the computed structure factor, $S(q)$, of the annealed structure for each model.
The \texttt{PaiNN} model provides excellent agreement with both its teacher counterpart and the general-purpose \texttt{ACE} potential.
We note that when aggressively modelling pores of large surface area by probing the very low density regimes ($<$ 1 g/cm$^3$), the \texttt{PaiNN} model is pushed beyond its domain and the MD simulations become unphysical.

\subsection*{Methylammonium lead iodide} 

Hybrid halide perovskites emerged in the early 2010s as promising materials for thin-film photovoltaics.\cite{lee_efficient_2012} 
The canonical representative, methylammonium lead iodide (MAPI), has since been the focus of many experimental\cite{whitfield_structures_2016, lopez_enhanced_2020} and computational\cite{eames_ionic_2015,wang_first-principles_2018} studies aimed at understanding its dynamic structure and electronic properties. 
While MAPI can adopt different polymorphs depending on temperature, the cubic form is the most relevant for the operation of solar-cell devices.\cite{ong_structural_2015}

We fine-tuned \texttt{MACE-MP-0b3} on 25 MAPI structures taken from Ref.~\citenum{bokdam2023mixed}, labelled with the meta-GGA SCAN functional as described in Ref.~\citenum{fykouras_disorder_2023}, before distilling into a \texttt{PaiNN} model using 5k synthetic structures. Both the teacher and the student model perform well on the DFT-labelled test set (29 and 32 meV/\AA\ force-component RMSEs, respectively).
For further validation, we ran constant-pressure MD simulations (1 ns, 350 K) on a $3 \times 3 \times 3$ cubic supercell.
The equilibrium pseudocubic lattice parameters during the 350 K run ranged between 6.3 and 6.4 \r{A} for both the fine-tuned FM and the distilled model, in accord with experimental\cite{nakamura_thermal_2025} and computational\cite{jinnouchi_phase_2019} studies at this temperature.

To compare teacher and student models in a more nuanced way, we calculate the molecular cation directions (along the C--N bond) and project these into spherical polar ($\theta$) and azimuthal ($\phi$) angles of the spherical coordinate systems.
We plot a distribution of these angles in Fig.~\ref{fig:applications}c: the foundation and distilled models generate near identical distributions, which are consistent with the nearly isotropic rotational dynamics of the molecular cation reported for the cubic phase of MAPI and related hybrid perovskite systems. \cite{mattoni_methylammonium_2015, fykouras_disorder_2023, pdyna}

\subsection*{An organic reaction in explicit solvent}

Finally, we explored an example of chemical reactivity: the prototypical second-order nucleophilic substitution (S$_{\text{N}}$2) reaction between 2-chloropropane and iodide in acetone.\cite{Conant_1925, Clayden_organic_chemistry} Despite its apparent simplicity, this reaction is challenging for modelling because there are competitive reaction pathways, and these are strongly affected by the reaction conditions. For example, primary (tertiary) alkyl halides typically show a clear S$_{\text{N}}$2 (S$_{\text{N}}$1) preference, respectively --- but secondary alkyl halides can undergo either pathway, with the dominant mechanism dictated by the nature of the nucleophile and the solvent. In the presence of a strong base, elimination can also become the favoured pathway. 

As a result of this complexity, our initial attempt to fine-tune the \texttt{MACE-OFF23} FM with explicitly solvated reactant state (RS), transition state (TS), and product state (PS) structures along the intrinsic reaction coordinate for the S$_{\text{N}}$2 mechanism was unsuccessful: the fine-tuned \texttt{MACE-OFF23} model over-stabilised the carbocation structure. Consequently, we employed active learning (AL) with well-tempered metadynamics sampling\cite{Zhang-24-07,Vitartas_2025_metadynamics} to generate a broader range of reactive configurations in both explicit solvent (26 acetone molecules, 101 structures) and the gas phase (214 structures). We randomly selected 150 of these structures, 75 from each sets, and used these to fine-tune \texttt{MACE-OFF24} (see Methods for details), giving final MAE values on the hold-out test set of 0.7 meV/atom and 18 meV/\AA{} for energies and forces, respectively.

We tested the stability of the resulting model through metadynamics simulation of the reactants solvated in a box with 48 acetone molecules. The fine-tuned \texttt{MACE-OFF24} model accurately predicted the concerted S$_{\text{N}}$2 mechanism. However, after 800 ps, dynamics yielded unphysical structures, presumably resulting from an inaccurate description of the repulsive interaction between alkyl and halogen atoms. Despite these limitations, we distilled the FM into a \texttt{PaiNN} model with 20k parameters using 5k synthetic structures generated with \texttt{augment-atoms}. The MAE of the student model on the DFT-labelled test set (1.5 meV/atom for energies and 29 meV/\AA{} for forces) was comparable to that of the teacher model. 

Metadynamics simulations driven by the distilled model reproduced the S$_{\text{N}}$2 mechanism (Fig~\ref{fig:applications}d). However, despite the potential remaining stable, unphysical reaction products were again observed after 600ps of simulation. We attribute this behaviour to the limitations of the teacher model in describing high-energy structures generated during metadynamics, including Cl--I pairs and structures with both halogens simultaneously too close to the alkyl. 
In summary, our study confirms that while distillation shows promise in generating fast and accurate reactive MLIPs for reactions in explicit solvent, the teacher and student models used here still suffer from the inaccurate description of strong attractive and repulsive interactions involved in the bond-breaking / -forming process, which are outside of the teacher's training set.

\section*{Discussion}

We have presented a general approach for the architecture-agnostic distillation of atomistic foundation models to target specific chemical domains. 
Our approach enables the creation of `student' MLIPs of any architecture that maintain the predictive power of their foundation model `teachers' while drastically reducing computational requirements: 
the distilled models can be substantially faster and less memory-intensive at inference time, making them valuable for large-scale atomistic simulations.
We have shown example applications across various domains of atomistic modelling that can be explored with atomistic FMs. 

Simultaneously, the distilled models obtained with our approach are quick and cheap to produce: the only human input required is a small number of ($<50$) DFT-labelled structures to fine-tune an existing FM to the desired level of accuracy.
Once the user provides their input structures, all steps in the pipeline are fully automated, implemented in open-sourced code, and benefit from GPU acceleration.
As such, the distillation process can take as little as 3 hours on a single Nvidia A6000 GPU. 
Together, the conceptual advantages of model distillation and the implementation presented here can provide improved access to accurate, performant, and robust MLIPs.
Given the substantial computational effort required for very-large-scale atomistic simulations, and the associated energy requirements and (indirectly) \ce{CO2} emissions, providing access to cheaper models where possible appears to be highly timely.

Looking forward, we envisage continued FM development (through improved architectures, fine-tuning methodologies, and larger datasets), as well as more accurate, smaller, and therefore faster MLIP architectures and models.
Our distillation approach can benefit from advances in both fields, and is expected to remain relevant in the longer term due to the archi\-tec\-ture-agnostic nature of our findings.
As FMs improve further, the discrepancy between ground-truth data (typically, DFT) and the FM predictions will decrease, yielding higher-quality synthetic data labels; improvements in model efficiency will enable access to longer and larger-scale simulations to study complex chemical processes.
In this way, distillation may be viewed as another major step --- following foundational work showing accuracy and data scalability --- in creating truly mainstream MLIPs for scientific research. 

\clearpage
\setstretch{1.1}

\section*{Methods}

\textbf{Fine-tuning atomistic foundation models.}
Current FMs are able to drive stable MD simulations that are qualitatively correct in a zero-shot setting,\cite{Batatia-24-03} but not necessarily in quantitative agreement with direct quantum-mechanical computations at the ground-truth level. 
Using only very small amounts of data, such FMs can be fine-tuned to reproduce ground-truth energies and forces with high accuracy for specific chemical systems.\cite{Yang-24-05, Kaur-25-01, kong_mattertune_2025} 
Beyond a simple continuation of training, several advanced protocols have been proposed for this purpose, including random feature selection\cite{Novelli-24}, frozen transfer learning, \cite{Radova-25-02} and low-rank adaptations. \cite{Mazitov-25-03}

Here, we used \texttt{graph-pes} \cite{graph-pes} to fine-tune the openly available \texttt{MatterSim-v1.0.0-1M}\cite{Yang-24-05}, \texttt{MACE-MP-0b3}\cite{Batatia-24-03}, \texttt{MACE-OFF24},\cite{Kovacs_JACS_MACEOFF_2025} and \texttt{orb-v3-direct-20-omat}\cite{Rhodes-25-04} FMs, as well as using the \texttt{graph-pes} interfaces to the \texttt{ASE}\cite{HjorthLarsen-17-06}, \texttt{LAMMPS},\cite{Thompson-22-02} and \texttt{torch-sim}\cite{Gangan-Torch-Sim-25} packages to run MD simulations.
We fine-tuned FMs using a fresh \texttt{Adam} optimiser as implemented in \texttt{PyTorch}, with an initial learning rate of $10^{-3}$ that decays by a factor of 0.8 after 25 epochs of no improvement on the validation set. Unless otherwise stated, we fine-tuned on 25 domain-specific structures, using a validation set containing a further 5 structures, and a batch size of 2. We found the learning curves during fine-tuning to be extremely smooth, and insensitive to the exact hyperparameter choices presented here, providing the learning rate did not exceed $3\times10^{-3}$.

\textbf{Synthetic data generation.}
To generate a large, synthetic dataset of atomistic structures, we take the small, existing dataset used to fine-tune the FM, and use the rattle-relax-repeat protocol detailed below to iteratively augment each structure separately.
Each of these new structures is then labelled with the fine-tuned FM (not DFT).
There is no information sharing between the augmentations of different structures, and so this process can be performed in a trivially parallel way for all starting structures.

For a given starting structure, our augmentation protocol constructs a `family tree' of atomic configurations; each step involves selecting a structure from this tree (\textit{i.e.}, a `parent'), rattling the atomic positions and unit cell, relaxing to get a new structure, and inserting this new structure (\textit{i.e.}, a `child') into the tree.
To begin with, the family tree consists of the single starting structure.

To choose a new parent structure, we randomly sample from all structures in the tree, where structure $i$ has a probability of being sampled given by
$$
\mathbb{P}_i = \beta \cdot \frac{e^{-E_i / kT}}{\sum_j e^{-E_j / kT}} + (1-\beta) \cdot \frac{G_i}{\sum_j G_j},
$$
where $E_i$ is the energy of structure $i$; $G_i \in \mathbb{Z}^+$ is the `generation' of the structure; and $\beta$ is a hyper-parameter controlling the explore-vs-exploit trade-off. 

To create a `child' from this parent structure, we perform the following transformation:
$$
R^\prime \leftarrow [(A + I) \times R] + B \; ; \qquad C^\prime \leftarrow (A + I) \times C_0,
$$
where $R$ are the atomic positions; $C_0$ is the unit cell of the original seed structure; $A \in \mathbb{R}^{3\times 3}$ has entries sampled from $\mathcal{N}(0, \sigma_{A})$; $B \in \mathbb{R}^{N \times 3}$ has entries sampled from $\mathcal{N}(0, \sigma_{B})$; and $\sigma_{A}$ and $\sigma_{B}$ are user-defined standard deviations controlling the extent of the rattling in the cell and positions, respectively. (In the case of isolated structures, we only rattle the positions, with $A = 0^{3 \times 3}$).

To relax this rattled child structure, we use energies and forces generated by the \textit{foundation model} (i.e., not DFT!) using a scheme inspired by the Robbins--Monro algorithm.\cite{Robbins-51-09}
Step $x$ of this relaxation involves updating the atomic positions according to
$$
R^\prime \leftarrow R + \frac{\sigma_B}{x} \cdot \frac{F}{||F||},
$$
where $F/||F||$ are the normalised unit vectors corresponding to the direction of each atomic force.
We perform up to $M$ relaxation steps, but stop early with probability $\min(0.25, e^{-\Delta E / kT})$ provided the maximum force magnitude is less than 30 eV/\AA{}, where $\Delta E$ is the energy difference between the relaxed child and its starting parent structure.

\begin{figure}[ht]
   \centering
   \includegraphics[width=\linewidth]{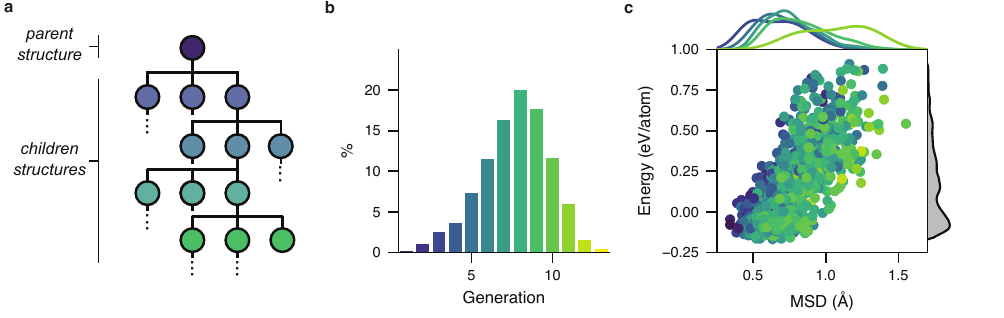} 
   \caption{
   \textbf{Synthetic data generation.} 
   (\textbf{a}) Our protocol creates a `family tree' of structures for each seed structure: in all panels, we colour-code each structure based on its generation.
   By incorporating an element of exploration into our structure selection step, we preferentially sample structures from later generations. The minimum distance between corresponding atoms in any two structures in the tree is at least 0.2 \AA.
   (\textbf{b}) We characterise the synthetic dataset of 1,000 structures generated from a single seed configuration of 64 molecules of water at ambient conditions; the percentage of structures for each generation number is shown.
   (\textbf{c}) MSD with respect to the seed (i.e., first-generation) structure against energy, colour-coded by generation number as in panels (a--b). The marginal distribution in energy shows a bias towards lower energies (exploitation), while still sampling higher ones. The marginal distribution in MSD shows that later generations move further away from the starting structure (exploration).}
   \label{fig:protocol}
\end{figure}

We perform this iterative rattle-relax-repeat procedure for each structure in the dataset used to fine-tune the model. We analyse a collection of 1,000 bulk water structures generated using this procedure in Fig.~\ref{fig:protocol}.
The protocol is implemented in the GPU-accelerated \texttt{augment-atoms} package and openly available at \url{https://github.com/jla-gardner/augment-atoms}.

Unless otherwise stated, we also generate a collection of dimer structures by uniformly sampling the interatomic distance in the interval $[0.6d, 3d]$, where $d$ is the sum of the relevant covalent radii, to create 50 structures for each unique pair of elements.

\textbf{Student models.}
We used the \texttt{PaiNN}\cite{Schutt-21-06} and \texttt{TensorNet}\cite{Simeon-23-06} architectures to create GNN-based `student' MLIPs.
As for the fine-tuning task described above, we again used the \texttt{graph-pes} package, now to directly train these models on our synthetic datasets.
We used the same \texttt{Adam} optimiser and learning-rate settings as for fine-tuning FMs.
Unless otherwise stated, we incorporated a ZBL repulsion term to help to model the repulsion of inner-shell electrons,\cite{Ziegler-85} and chose hyperparameters that yield fast models with $\sim$40k parameters.
(In the present study, we used a separate batch size for each system, to maximise the utilisation of the GPU.)

For ACE models, we optimised hyperparameters using \texttt{XPOT}\cite{ThomasduToit_TJoCP_Crossplatform_2023} on a dataset of 333 synthetic structures.
50 iterations were performed, and the hyperparameters of the model with the lowest loss on a validation set were used to fit the models used in this paper.
All ACE models were trained with \texttt{pacemaker}\cite{Lysogorskiy-21-06} on an Nvidia RTX A4500 GPU.
We used a fresh \texttt{AdamW} optimiser with a batch size of 32, $\kappa = 0.75$, and an automatically determined ZBL repulsion term, similar to MLIPs trained using \texttt{graph-pes}.
If the loss on the validation set did not improve by $1\times10^{-5}$ per epoch over any 50 epochs, the potential stopped fitting.
Finally, BFGS optimisation was used to `upfit' (refine) the model for 50 epochs with $\kappa = 0.01$ to improve energy-prediction accuracy.

EDDP models were fitted using the reference implementation described in Refs.\ \citenum{Pickard-22-07} and \citenum{salzbrenner_developments_2023}. Specifically, 256 individual neural-network (two layers of 20 nodes) potentials were trained on 80\% of the data with randomly initialised weights. These potentials were combined (ensembled) using a non-negative least squares fit to 10\% of the data, a validation set which is also used to early-stop the Levenberg--Marquadt training. The remaining 10\% of the data were used to estimate the test error, separately to the error reported in Fig \ref{fig:ablations}a. The feature vectors were constructed with exponents ranging from 2 to 10, with 16 and $4^2$ for the two and three-body terms respectively, along with a cut-off radius of 5.5 \AA{}.

\textbf{Applications to materials.}
For high-pressure hydrogen, we ran a series of MD simulations, similar to the ones in Ref.~\citenum{BenMahmoud-22-09}, using the i-PI~\cite{Litman-24-08} interface with LAMMPS~\cite{Thompson-22-02}, targeting the NPT ensemble for ionic temperatures of 10,000, 20,000, 35,000, and 50,000~K, and pressures ranging from 400~GPa to 1,600~GPa. All simulations ran for 25~ps, with the exception of a few trajectories in the higher density/pressure regime. In particular, the trajectory at 1,500~GPa and 50,000~K failed after 5~ps, that at 1,600~GPa and 35,000~K after 10~ps, and that at 1,600~GPa and 50,000~K after 2~ps. These simulations failed despite reducing the timestep to 0.001~fs.

For silica, we ran MD to generate a-\ce{SiO2} starting configurations  using a general-purpose \texttt{ACE} model for the Si--O system \cite{Erhard-24-03} and a fast quench rate of  10$^{14}$ K/s. The simulations were run at constant cell volume (NVT ensemble) to maintain the low density of 1.2 g/cm$^3$, and resulted in non-equilibrated, mesoporous a-\ce{SiO2} structures of 1,200 atoms. We then compared our \texttt{PaiNN} model to the \texttt{orb-v3-FT} and Si--O \texttt{ACE} models by running 10 ps of MD with ASE's Berendsen NPT implementation, using a timestep of 1 fs at 500 K and 1 bar. For each model, this procedure was repeated four times from different starting configurations.

For MAPI, we ran MD simulations using Nos\'e--Hoover--Parinello--Rahman NPT dynamics\cite{Melchionna_MP_Hoover_1993, Melchionna_PRE_Constrained_2000} as implemented in ASE\cite{HjorthLarsen-17-06}, with a timestep of 1 fs. For projecting the C--N vectors onto spherical coordinates, we adapted the method described in Ref.~\citenum{pdyna}.

\textbf{Organic reaction in solution.}
Quantum-chemical computations were performed with Orca 6.0.0\cite{orca_neese_2022} at the $\omega$B97M-D3BJ/def2-TZVP level of theory.\cite{Goerigk_omega-dft_2018} Active learning was performed with the \texttt{mlp-train} package,\cite{Young_PCCP_Reaction_2022} using well-tempered MetaD sampling with inherited bias combined with a SOAP-based similarity selector. The `small' \texttt{MACE-OFF23} model\cite{Kovacs_JACS_MACEOFF_2025} was used as the initial MLIP in active learning and was fine-tuned before each loop. For the production runs, we fine-tuned the `medium' \texttt{MACE-OFF24} model\cite{Kovacs_JACS_MACEOFF_2025}, using a custom loss function that assigned higher weights to the forces on the reactive species (25 for substrate C, Cl and I atoms; 10 for substrate H) compared to those on the solvent (weight of 1).

Production well-tempered MetaD simulations were done in i-PI 3.1.2,\cite{Litman-24-08} coupled to Plumed 2.9.2.\cite{plumed2_2014} The simulation ran with a 0.5 fs time step in the NVT ensemble at 323 K, using a stochastic velocity rescaling thermostat with a coupling time constant of 100 fs.\cite{Bussi_rescaling_2007} The coordination numbers between the central C and Cl / I atoms, respectively, served as collective variables with $R_0=0.8$ and $D_0$ equal to 1.8 (2.0) for Cl (I), respectively. The  MetaD simulations used Gaussians with widths set to 0.1 for both CVs and height 0.1 kcal/mol, which were deposited every 25 fs with a bias factor of 20.  The maximum value of the C--Cl/I distance was restrained by upper wall bias at 5 \AA{} with a force constant of 1,000 kcal/mol/\AA{}.

\section*{Data availability}

Data supporting this work are available at \href{https://github.com/dft-dutoit/synthetic-distillation}{https://github.com/dft-dutoit/synthetic-distillation}.

\section*{Code availability}

The data generation pipeline described herein is available as an open-sourced Python implementation at \href{https://github.com/jla-gardner/augment-atoms}{https://github.com/jla-gardner/augment-atoms} under the MIT license. 
The {\tt graph-pes} software, used to train and fine-tune graph-based MLIP models, is available at \href{https://github.com/jla-gardner/graph-pes}{https://github.com/jla-gardner/graph-pes} under the MIT licence. 
The {\tt XPOT} software, used for hyperparameter optimisation for ACE models, is available at \href{https://github.com/dft-dutoit/xpot}{https://github.com/dft-dutoit/xpot} under the GPL-2.0 licence.
Other software was used as provided by their respective authors.

\section*{Acknowledgements}
J.L.A.G. acknowledges a UKRI Linacre - The EPA Cephalosporin Scholarship, support from an EPSRC DTP award [grant number EP/T517811/1], and from the Department of Chemistry, University of Oxford.
C.B.M. acknowledges funding from the Swiss National Science Foundation (SNSF) under grant number 217837.
Z.F.B. was supported through an Engineering and Physical Sciences Research Council DTP award and IBM Research.
V.J. acknowledges the funding from Schmidt Sciences, LLC.
L.-B.P. acknowledges funding from the EPSRC Centre for Doctoral Training in Inorganic Chemistry for Future Manufacturing (OxICFM), EP/S023828/1.
This work was supported by UK Research and Innovation [grant number EP/X016188/1].
We are grateful for computational support from the UK national high performance computing service, ARCHER2, for which access was obtained via the UKCP consortium and funded by EPSRC grant ref EP/X035891/1.
The authors would like to acknowledge the use of the University of Oxford Advanced Research Computing (ARC) facility in carrying out this work (http://dx.doi.org/10.5281/zenodo.22558).

\section*{Author contributions}
J.L.A.G. and D.F.T.d.T. conceived the study. J.L.A.G., D.F.T.d.T., and V.L.D. designed and coordinated the project.
J.L.A.G. conceived the synthetic data generation protocol, fine-tuned foundation models, generated synthetic data, and trained the graph-network models;
D.F.T.d.T. trained all ACE models;
C.J.P. trained all EDDP models.
Z.F.B. and F.M. validated MLIP models for water.
Applications were contributed by C.B.M. (high-pressure hydrogen), L.A.M.R. (silica), L.-B.P. (MAPI), and V.J. and F.D. (organic molecules), respectively.
J.L.A.G., D.F.T.d.T., C.J.P., and V.L.D. developed the central conclusions. 
J.L.A.G., D.F.T.d.T., and V.L.D. wrote the paper with input from all authors.

\end{document}